\documentclass[aps,prl,twocolumn,superscriptaddress,nofootinbib]{revtex4-2}

\usepackage{amsmath,amssymb,bm,graphicx}
\usepackage[colorlinks=true,citecolor=blue,linkcolor=blue,urlcolor=blue]{hyperref}

\newcommand{\kvec}{\mathbf{k}}
\newcommand{\pvec}{\mathbf{p}}
\newcommand{\qvec}{\mathbf{q}}

\newcommand{\cD}{\mathcal{D}}
\newcommand{\cH}{\mathcal{H}}
\newcommand{\cP}{\mathcal{P}}
\newcommand{\cI}{\mathcal{I}}
\newcommand{\cF}{\mathcal{F}}
\newcommand{\Tr}{\operatorname{Tr}}
\newcommand{\Cov}{\operatorname{Cov}}

\begin{document}
	
	\title{Quantum Memory in Scalar-Induced Gravitational Waves}
	
	\author{Waqas Ahmed}
	\affiliation{
		Center for Fundamental Physics and School of Artificial Intelligence,
		Hubei Polytechnic University, Huangshi 435003, China
	}
	
	\begin{abstract}
		Scalar-induced gravitational waves are usually treated as a classical stochastic
		background sourced by phase-random curvature perturbations. We show that this
		description can miss residual quantum information. Starting from a decohered
		two-mode Gaussian scalar state, we derive explicit transfer relations between
		the scalar anomalous coherence and the covariance matrix of induced tensor
		modes. For a localized scalar power spectrum, the ordinary tensor power is
		sourced by scalar power contractions, whereas the opposite-mode tensor
		coherence is sourced by scalar anomalous-coherence contractions. This coherence
		can generate nonzero tensor discord and a connected tensor-power covariance
		even after scalar entanglement has vanished. We identify the connected
		covariance and phase-sensitive strain correlations as probes of primordial
		quantum coherence in secondary gravitational-wave backgrounds, and discuss
		their possible relevance for future space-based interferometers and pulsar
		timing arrays.
	\end{abstract}
	
	\maketitle
	
	
	\textit{Introduction.---}
	Inflation provides a quantum origin for the primordial density perturbations
	that seed cosmic structure. Vacuum fluctuations are amplified and squeezed on
	super-Hubble scales, producing strongly correlated pairs of Fourier modes
	$(\kvec,-\kvec)$~\cite{Guth:1980zm,Linde:1981mu,Mukhanov:1990me,Polarski:1995jg}.
	A long-standing question is whether any genuinely quantum feature of these
	perturbations can survive the quantum-to-classical transition and appear in
	late-time observables~\cite{Kiefer:1998qe,Schlosshauer:2014pgr,Zurek:2003zz}.
	
	Most discussions of primordial quantumness focus on entanglement. However,
	entanglement is fragile: environmental decoherence can rapidly suppress it,
	especially for highly squeezed cosmological perturbations. Quantum discord
	provides a broader diagnostic of nonclassical correlations. Unlike
	entanglement, discord can remain finite in mixed separable states
	\cite{Henderson:2001wrr,Modi:2012baj}. This makes it a natural candidate for
	residual quantum information in the primordial Universe.
	
	Scalar-induced gravitational waves (SIGWs) provide a natural arena for testing
	this idea. SIGWs are generated at second order when scalar perturbations source
	tensor modes through nonlinear gravitational interactions
	\cite{Ananda:2006af,Baumann:2007zm,Saito:2008jc,Domenech:2021ztg}. Since the
	source is quadratic in the curvature perturbation, the induced tensor state is
	sensitive not only to the scalar power spectrum, but also to phase-space
	correlations encoded in the scalar covariance matrix.
	
	In this Letter we show that residual scalar coherence can be transferred to
	induced tensor modes. We use a Gaussian covariance-matrix description of
	opposite-momentum mode pairs and derive explicit transfer relations linking
	the scalar anomalous coherence to the induced tensor covariance. The resulting
	tensor coherence generates tensor discord, connected tensor-power covariance,
	and phase-sensitive strain correlations. These observables probe information
	beyond the ordinary stochastic power spectrum.
	
	
	\textit{Scalar discord after decoherence.---}
	For a normalized pair of scalar modes $(\kvec,-\kvec)$, define the quadrature
	vector
	\begin{equation}
		\hat{\bm{\xi}}_{\zeta}
		=
		\left(
		\hat q_{\kvec},\hat p_{\kvec},
		\hat q_{-\kvec},\hat p_{-\kvec}
		\right)^T .
	\end{equation}
	A two-mode squeezed scalar state has covariance matrix
	\begin{equation}
		\sigma_{\rm sq}
		=
		\frac12
		\begin{pmatrix}
			A & C\\
			C^T & A
		\end{pmatrix},
		\qquad
		A=\cosh(2r_k)\mathbb I_2 ,
		\label{eq:sq_covariance}
	\end{equation}
	with
	\begin{equation}
		C=
		\sinh(2r_k)
		\begin{pmatrix}
			\cos\theta_k & \sin\theta_k\\
			\sin\theta_k & -\cos\theta_k
		\end{pmatrix}.
		\label{eq:sq_c_block}
	\end{equation}
	Here $r_k$ is the squeezing amplitude and $\theta_k$ is the squeezing angle.
	The off-diagonal block $C$ encodes correlations between the modes
	$\kvec$ and $-\kvec$.
	
	To model environmental decoherence while keeping analytic control, we add an
	isotropic Gaussian-noise channel,
	\begin{equation}
		\sigma_{\zeta}
		=
		\sigma_{\rm sq}
		+
		\epsilon_{\rm dec}\mathbb I_4 ,
		\label{eq:scalar_noise}
	\end{equation}
	where $\epsilon_{\rm dec}$ is the decoherence strength. This parameter should
	be understood as an effective description of the cumulative influence of
	unobserved environmental degrees of freedom, such as sub-Hubble modes,
	additional light fields, or higher-order gravitational interactions. The
	channel preserves Gaussianity and is physical for $\epsilon_{\rm dec}\geq0$.
	Indeed, for $\theta_k=0$ the symplectic eigenvalues are
	\begin{equation}
		\nu_{\pm}
		=
		\sqrt{
			\frac14
			+
			\epsilon_{\rm dec}\cosh(2r_k)
			+
			\epsilon_{\rm dec}^2
		}
		\geq
		\frac12 ,
		\label{eq:physicality_noise}
	\end{equation}
	so the uncertainty condition $\sigma_\zeta+i\Omega/2\geq0$ is satisfied.
	
	For phase-insensitive quantities one may set $\theta_k=0$ and define
	\begin{equation}
		a_k=\frac12\cosh(2r_k)+\epsilon_{\rm dec},
		\qquad
		c_k=\frac12\sinh(2r_k).
		\label{eq:ak_ck}
	\end{equation}
	The smallest symplectic eigenvalue of the partially transposed covariance
	matrix is then
	\begin{equation}
		\widetilde{\nu}_{-}
		=
		a_k-c_k
		=
		\frac12 e^{-2r_k}+\epsilon_{\rm dec}.
		\label{eq:nu_tilde}
	\end{equation}
	The scalar state is entangled only when $\widetilde{\nu}_{-}<1/2$. Thus
	sufficient decoherence removes scalar entanglement. Gaussian discord, however,
	may remain finite in part of the separable regime. This hierarchy motivates
	using discord and anomalous coherence as probes of residual quantum
	information.
	
	The scalar anomalous-coherence fraction entering the tensor source is
	\begin{equation}
		\chi_{\rm dec}(k)
		=
		\frac{c_k}{a_k}
		=
		\frac{
			\frac12\sinh(2r_k)
		}{
			\frac12\cosh(2r_k)+\epsilon_{\rm dec}
		} .
		\label{eq:chi_dec}
	\end{equation}
	This quantity is close to unity for a highly coherent squeezed state and is
	suppressed by decoherence. Importantly, it does not vanish at the same point
	at which the logarithmic negativity becomes zero.
	
	
	\textit{Explicit scalar-to-tensor transfer.---}
	The induced tensor mode obeys
	\begin{equation}
		h_{\kvec}^{\lambda\prime\prime}
		+
		2\cH h_{\kvec}^{\lambda\prime}
		+
		k^2h_{\kvec}^{\lambda}
		=
		S_{\kvec}^{\lambda},
		\label{eq:tensor_eom}
	\end{equation}
	where $\lambda=+,\times$. The scalar source is quadratic,
	\begin{equation}
		S_{\kvec}^{\lambda}(\tau)
		=
		4
		\int
		\frac{d^3p}{(2\pi)^3}
		e_{ij}^{\lambda}(\kvec)p^ip^j\,
		\cF(p,q,\tau)
		\zeta_{\pvec}\zeta_{\kvec-\pvec},
		\label{eq:source}
	\end{equation}
	with $q=|\kvec-\pvec|$. The late-time induced tensor operator is
	\begin{equation}
		\hat h_{\kvec}^{\lambda}
		=
		\int_{\tau_i}^{\tau_{\rm obs}}
		d\tau\,
		G_k(\tau_{\rm obs},\tau)
		\hat S_{\kvec}^{\lambda}(\tau),
		\label{eq:green_solution}
	\end{equation}
	where $G_k$ is the retarded Green function.
	
	We work to leading nontrivial order in the scalar perturbations. Since the
	induced tensor field is sourced as $h_{\kvec}\sim\zeta^2$, the tensor
	two-point covariance is of order $\zeta^4$. For a Gaussian scalar state, this
	order is completely determined by scalar four-point functions, which factorize
	through Wick contractions. Higher-order corrections would involve scalar
	six-point and higher correlators and are neglected in the present Letter.
	
	We take the scalar power spectrum to be localized,
	\begin{equation}
		\cP_\zeta(k)
		=
		A_\zeta
		\exp
		\left[
		-\frac{\ln^2(k/k_*)}{2\sigma_\zeta^2}
		\right],
		\label{eq:scalar_lognormal}
	\end{equation}
	and parametrize the scalar anomalous coherence as
	\begin{equation}
		\mathcal C_\zeta(k)
		=
		\chi_{\rm dec}(k)\,
		\cP_\zeta(k)\,
		e^{i\theta_k}.
		\label{eq:scalar_coherence}
	\end{equation}
	The important point is that $\chi_{\rm dec}$ is fixed by the scalar covariance
	matrix in Eq.~\eqref{eq:chi_dec}; it is not introduced as an independent
	tensor parameter.
	
	The tensor reduced state is obtained from
	\begin{equation}
		\rho_h
		=
		\Tr_{\zeta}
		\left[
		U\rho_0U^\dagger
		\right],
		\qquad
		\rho_0
		=
		|0\rangle_h\langle0|_h
		\otimes
		\rho_{\zeta}.
		\label{eq:rhoh}
	\end{equation}
	At leading order the late-time tensor pair $(\kvec,-\kvec)$ may be described
	by an effective Gaussian covariance matrix. In a phase convention where the
	opposite-mode coherence is real, it takes the form
	\begin{equation}
		\sigma_h(k)
		=
		\begin{pmatrix}
			\alpha_k & 0 & \gamma_k & 0\\
			0 & \alpha_k & 0 & -\gamma_k\\
			\gamma_k & 0 & \alpha_k & 0\\
			0 & -\gamma_k & 0 & \alpha_k
		\end{pmatrix},
		\label{eq:tensor_cm}
	\end{equation}
	where $\alpha_k=1/2+d_k$. The diagonal tensor occupation is
	\begin{equation}
		d_k
		=
		\int d\Pi_{\pvec}\,
		\left|
		\cI(k,p,q)
		\right|^2
		\cP_\zeta(p)\cP_\zeta(q),
		\label{eq:dk_integral}
	\end{equation}
	where $d\Pi_{\pvec}$ denotes the induced-gravitational-wave phase-space
	measure and $\cI(k,p,q)$ is the time-integrated source kernel. The
	opposite-mode tensor coherence is
	\begin{equation}
		\gamma_k
		=
		\int d\Pi_{\pvec}\,
		\cI(k,p,q)^2\,
		\mathcal C_\zeta(p)\mathcal C_\zeta(q).
		\label{eq:gammak_integral}
	\end{equation}
	Equations~\eqref{eq:dk_integral} and \eqref{eq:gammak_integral} are the central
	transfer relations. The ordinary tensor power is generated by scalar power
	contractions, whereas the tensor coherence is generated by scalar
	anomalous-coherence contractions. 
	For notational simplicity we have written the parity-symmetric form of the
	kernel product. In the most general case, $\mathcal I(k,p,q)^2$ should be
	replaced by $\mathcal I(k,p,q)\mathcal I(-k,-p,-q)$.

	For a narrow scalar spectrum, the integrals are dominated by
	$p\simeq q\simeq k_*$ and have support mainly for $k\lesssim2k_*$. One obtains
	the transparent scaling
	\begin{equation}
		d_k
		\simeq
		K_A(k)\,A_\zeta^2,
		\qquad
		\gamma_k
		\simeq
		K_\gamma(k)\,
		\chi_{\rm dec}^2
		A_\zeta^2
		e^{2i\theta_*},
		\label{eq:narrow_scaling}
	\end{equation}
	where $K_A$ and $K_\gamma$ are dimensionless transfer kernels. This shows
	explicitly that residual scalar anomalous coherence produces residual tensor
	coherence.
	
	A classical stochastic field with imposed phase correlations could also
	generate a nonzero connected covariance. In the present case, however, the
	opposite-mode tensor coherence is not chosen independently: it is fixed by the
	scalar covariance matrix through $\chi_{\rm dec}$ and by the scalar-to-tensor
	transfer kernel. The connected covariance should therefore be interpreted as a
	probe of transferred opposite-mode coherence. Its quantum interpretation
	follows when the underlying scalar state has nonzero Gaussian discord.
	
	\begin{figure}[t]
		\centering
		\includegraphics[width=\columnwidth]{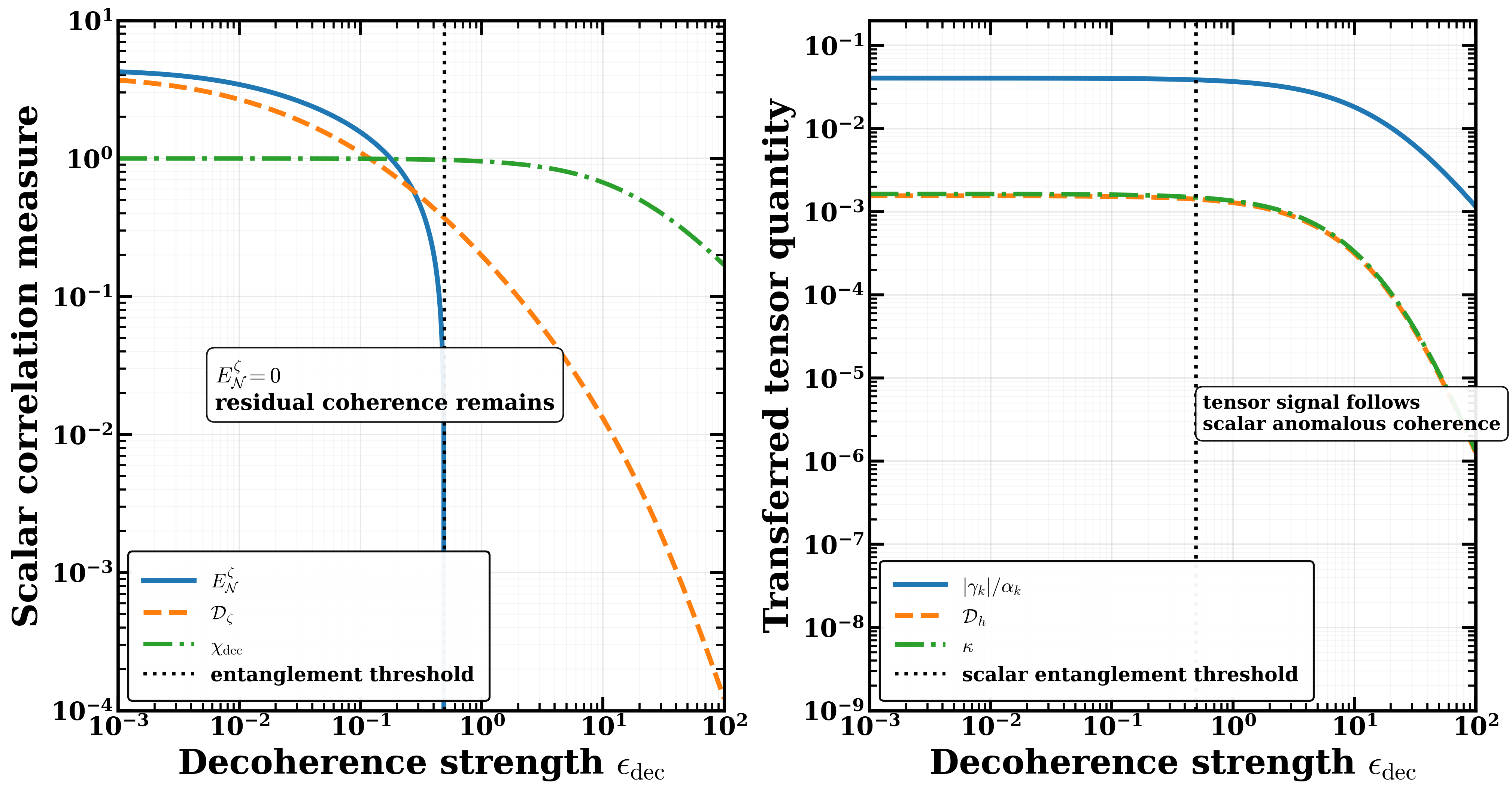}
		\caption{
			Minimal scalar-to-tensor transfer benchmark. Panel (left)  shows the scalar
			logarithmic negativity, Gaussian discord, and anomalous coherence fraction
			as functions of the decoherence strength. The vertical line marks the point
			where scalar entanglement vanishes. Panel (right) shows the induced tensor
			coherence, tensor discord, and connected covariance generated by the same
			residual scalar anomalous coherence. The tensor quantities remain nonzero
			inside the scalar separable regime because they are sourced by coherence
			contractions rather than by scalar entanglement itself.
		}
		\label{fig:transfer}
	\end{figure}
	
	
	\textit{Tensor discord.---}
	The Gaussian tensor discord is computed from the symplectic invariants of
	Eq.~\eqref{eq:tensor_cm}. Physicality requires
	\begin{equation}
		\alpha_k^2-|\gamma_k|^2\geq\frac14.
		\label{eq:tensor_physicality}
	\end{equation}
	The symplectic eigenvalue is
	\begin{equation}
		\nu_h
		=
		\sqrt{\alpha_k^2-|\gamma_k|^2}.
	\end{equation}
	For the symmetric two-mode state, the Gaussian discord is
	\begin{equation}
		\cD_h(k)
		=
		f(\alpha_k)
		-
		2f(\nu_h)
		+
		f
		\left(
		\alpha_k
		-
		\frac{\gamma_k^2}{\alpha_k+1/2}
		\right),
		\label{eq:tensor_discord}
	\end{equation}
	where
	\begin{equation}
		f(x)
		=
		\left(x+\frac12\right)\ln\left(x+\frac12\right)
		-
		\left(x-\frac12\right)\ln\left(x-\frac12\right).
	\end{equation}
	Writing $\alpha_k=1/2+d_k$, the weak-coherence expansion gives
	\begin{equation}
		\cD_h(k)
		=
		\frac{
			\ln[(1+d_k)/d_k]
		}{
			(1+d_k)(1+2d_k)
		}
		|\gamma_k|^2
		+
		\mathcal O(|\gamma_k|^4).
		\label{eq:discord_expansion}
	\end{equation}
	Together with Eq.~\eqref{eq:gammak_integral}, this demonstrates that tensor
	discord is generated by scalar anomalous coherence entering the second-order
	source. The precise mapping between scalar discord and tensor discord is
	kernel dependent, but the transfer mechanism is explicit.
	
	
	\textit{Connected covariance and phase-sensitive observables.---}
	The ordinary SIGW spectrum is controlled mainly by $d_k$. A more direct probe
	of the off-diagonal tensor covariance is the connected tensor-power statistic
	\begin{equation}
		\kappa(k)
		=
		\frac{
			\Cov
			\left(
			|\hat h_{\kvec}|^2,
			|\hat h_{-\kvec}|^2
			\right)
		}{
			\langle|\hat h_{\kvec}|^2\rangle
			\langle|\hat h_{-\kvec}|^2\rangle
		} .
		\label{eq:kappa_def}
	\end{equation}
	For a phase-random Gaussian reference background, the anomalous coherence
	vanishes and $\kappa=0$. For the Gaussian tensor state in
	Eq.~\eqref{eq:tensor_cm}, Wick's theorem gives
	\begin{equation}
		\kappa(k)
		\simeq
		c_\kappa
		\frac{|\gamma_k|^2}{\alpha_k^2},
		\qquad
		c_\kappa>0 ,
		\label{eq:kappa}
	\end{equation}
	where $c_\kappa$ depends on the precise normalization convention for the power
	estimator. Thus $\kappa(k)$ directly probes the off-diagonal tensor coherence.
	
	The tensor coherence may also carry a phase,
	\begin{equation}
		\gamma_k=|\gamma_k|e^{i\Theta_k}.
	\end{equation}
	In that case the real quadrature correlation block contains both
	${\rm Re}\,\gamma_k$ and ${\rm Im}\,\gamma_k$. Complex strain correlators,
	such as
	\begin{equation}
		\Gamma(\kvec,\qvec)
		=
		\langle
		\hat h_{\kvec}
		\hat h_{\qvec}
		\rangle ,
	\end{equation}
	can therefore contain phase information not captured by the power spectrum.
	Such phase-sensitive observables may help distinguish transferred tensor
	coherence from a purely phase-random stochastic background.
	
	\begin{figure}[t]
		\centering
		\includegraphics[width=\columnwidth]{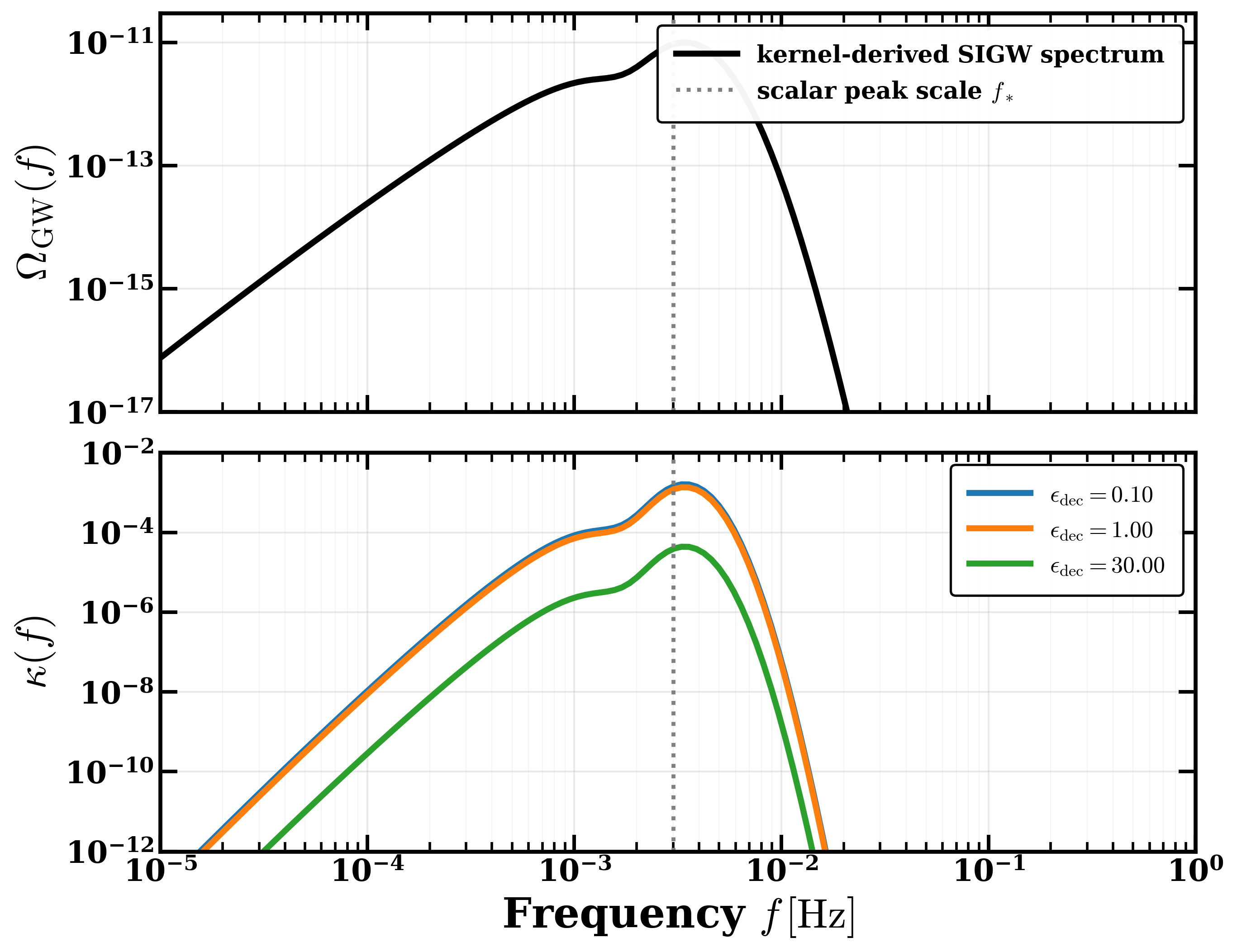}
		\caption{
			Kernel-derived benchmark signals from a localized scalar source. Panel (a)
			shows the scalar-induced gravitational-wave spectrum obtained from the
			radiation-era transfer kernel for a log-normal scalar power spectrum. The
			vertical line denotes the scalar peak scale $f_*$. Panel (b) shows the
			connected covariance statistic $\kappa(f)$ generated by the transferred
			tensor coherence for representative decoherence strengths. Both panels are
			computed from the same scalar spectrum and transfer kernel; the overall
			normalization is chosen for a representative detectable SIGW amplitude.
		}
		\label{fig:signals}
	\end{figure}
	
	
	\textit{Observational perspective.---}
	Equations~\eqref{eq:dk_integral}--\eqref{eq:kappa} provide the link between
	the scalar covariance matrix and tensor observables. A complete detector-level
	forecast would require the exact response functions of the detector network,
	foreground modeling, correlated-noise treatment, and a dedicated likelihood
	pipeline. Here we only give an order-of-magnitude scaling estimate for the
	connected covariance statistic. If $N_b$ approximately independent frequency
	bins contribute and the uncertainty is dominated by instrumental noise, the
	signal-to-noise ratio scales as
	\begin{equation}
		{\rm SNR}_{\kappa}
		\sim
		\sqrt{N_b}\,
		\kappa_0
		\left(
		\frac{\langle\Omega_{\rm GW}\rangle}
		{\Omega_{\rm noise}}
		\right)^2 .
		\label{eq:snr_kappa}
	\end{equation}
	This expression should not be interpreted as a full sensitivity curve; it only
	shows that a small connected signal can become relevant when the ordinary
	scalar-induced background is measured with high signal-to-noise ratio.
	
	Future space-based interferometers such as LISA, DECIGO, BBO, and TianQin are
	sensitive to the milli-Hertz to Hertz regime, while pulsar timing arrays probe
	the nano-Hertz band~\cite{LISA:2017pwj,NANOGrav:2023gor,EPTA:2023fyk,Xu:2023wog}.
	These frequency windows correspond to different scalar peak scales. The target
	observable identified here is not an arbitrary modulation of the power
	spectrum, but the connected covariance and phase structure generated by the
	transferred tensor coherence.
	
	
	\textit{Conclusions.---}
	We have shown that scalar-induced gravitational waves can carry information
	beyond the ordinary scalar power spectrum. Starting from a decohered Gaussian
	scalar state, we derived explicit transfer relations showing that scalar power
	contractions source the diagonal tensor covariance, while scalar anomalous
	coherence contractions source the opposite-mode tensor coherence. This
	coherence generates tensor discord and a connected covariance of tensor power,
	even in a regime where scalar entanglement has already vanished.
	
	The most robust signatures are therefore not universal shifts of the
	gravitational-wave spectrum, but observables sensitive to covariance and phase
	structure. The connected power covariance $\kappa(k)$ and phase-sensitive
	strain correlators provide natural targets for future searches. A full
	assessment of detectability will require detector-specific response functions,
	foreground modeling, and optimal four-point estimators. These results suggest
	that secondary gravitational-wave backgrounds may offer a new route to probing
	the quantum-information content of primordial perturbations.
	
	\bibliographystyle{apsrev4-2}
	\bibliography{PRL}
	
\end{document}